\documentclass{article}
\usepackage[T1]{fontenc}
\usepackage[utf8]{inputenc}
\usepackage{ismir} % Remove "submission" for camera-ready
\renewcommand{\permission}{}
\usepackage{amsmath,cite,url}
\usepackage{graphicx}
\usepackage{booktabs}
\usepackage{tikz}
\usetikzlibrary{positioning, arrows.meta, fit, calc}
\usepackage{stfloats} % enables [b] placement for table* and figure*

\graphicspath{{./}{../}}

\title{How Well Do Generative Music Models Follow Emotion Conditioning?}

\oneauthor
  {Morteza Heydari}
  {Istinye University, Graduate School, Istanbul, Turkey\\
  \texttt{coolmh75@gmail.com}}

\def\authorname{Morteza Heydari}

\begin{document}

\maketitle
\thispagestyle{empty}

\begin{abstract}
Recent generative music models offer increasingly fine-grained control through text and audio conditioning, yet how faithfully they follow intended emotional cues remains an open question. We address this gap with a unified evaluation pipeline for emotion-following in generated music. Using all 1000 tracks in GTZAN, we extract semantic audio descriptions with DashengLM, an audio captioning model, and estimate source-track valence and arousal with Music2Emotion, a music emotion recognition model. We construct affect-aware text prompts by combining descriptions with top-ranked emotion tags and generate 30-second outputs with three systems, Stable Audio Open, MusicGen, and InspireMusic, evaluating both text- and audio-conditioned generation. To measure emotion-following, we compute valence and arousal on generated audio and compare them with the source tracks using absolute error and Euclidean distance in valence–arousal space. Text-conditioned generation consistently outperforms audio conditioning, with MusicGen (text) and InspireMusic (text) achieving the best performance, while audio-conditioned variants prove less stable. We further find that valence is preserved more reliably than arousal and that emotion-following varies substantially across genres. These findings underscore the importance of evaluating affective controllability directly rather than relying solely on general quality or prompt-relevance metrics.
\end{abstract}

%% ================================================================
\section{Introduction}\label{sec:introduction}

Recent generative music models have substantially improved in realism, duration, and controllability.
While early deep-learning approaches targeted symbolic-domain generation with recurrent, adversarial, and variational architectures~\cite{WangEtAl2024_review_music_generation, YangEtAl2017_midinet}, modern systems operate directly in the audio domain and support conditioning through text, melody, or audio prompts.
Representative examples include MusicLM~\cite{AgostinelliEtAl2023_musiclm}, MusicGen~\cite{CopetEtAl2023_musicgen}, Stable Audio Open~\cite{EvansEtAl2025_stable_audio_open}, and InspireMusic~\cite{ZhangEtAl2025_inspiremusic}.

As these models become more capable, controllability becomes as important as audio quality.
Among controllable attributes, emotion is especially significant because music is fundamentally an affective medium.
In computational work, emotion is commonly represented through the valence--arousal model, where valence captures the positive--negative dimension and arousal captures the calm--energetic dimension~\cite{Russell1980_circumplex_affect}.

Despite the importance of affective control, systematic evaluation of emotion-following in generated music remains limited.
Recent surveys emphasize persistent challenges including limited emotion-labeled data, listener disagreement, and inconsistent evaluation protocols~\cite{DashAgres2024_affective_music_generation_review, LiyanarachchiJoshiMeijering2025_mmer_survey}.
Moreover, current evaluation practice in text-to-music systems primarily relies on Fr\'echet Audio Distance (FAD)~\cite{KilgourEtAl2019_fad}, KL-based distributional measures~\cite{KullbackLeibler1951_information_sufficiency}, CLAP-based text--audio relevance~\cite{ElizaldeEtAl2023_clap}, and human judgments of overall quality.
While valuable, none of these metrics directly quantify whether generated music matches an intended emotional profile.

We address this gap with a unified evaluation pipeline for emotion-following in generative music. Given all 1,000 tracks in GTZAN~\cite{tzanetakis2002gtzan}, we first extract a semantic audio description with DashengLM~\cite{MiLMPlus2025_midashenglm} and obtain categorical emotion tags along with continuous valence--arousal estimates with Music2Emotion~\cite{KangHerremans2025_music2emotion}. From these, we construct affect-aware text prompts that combine the audio description with the top-ranked emotion tags. We then generate music under two conditioning regimes---text conditioning using the constructed prompts, and audio conditioning using the original source tracks directly---and measure emotion-following by comparing valence--arousal estimates between each generated output and its corresponding source. Our contributions are:

\begin{enumerate}
    \item A reproducible evaluation pipeline for measuring emotion-following in music generation using continuous valence--arousal comparisons.
    \item A systematic comparison of five generation configurations across three models, revealing that text conditioning with explicit emotion cues significantly outperforms audio conditioning.
    \item Genre-wise and per-dimension analyses showing that valence is preserved more reliably than arousal and that emotion-following varies substantially across genres.
\end{enumerate}

%% ================================================================
\section{Related Work}\label{sec:related}

\subsection{Music Generation and Conditioning}

Automatic music generation has progressed from statistical models~\cite{Conklin2003_statistical} through symbolic neural generation with GANs and transformers~\cite{YangEtAl2017_midinet, HuangEtAl2019_music_transformer} to modern audio-domain systems.
Jukebox models raw audio through hierarchical discrete representations and autoregressive transformers~\cite{DhariwalEtAl2020_jukebox}, while diffusion-based approaches offer alternatives through iterative latent-space refinement~\cite{MittalEtAl2021_symbolic_diffusion}.
A major recent direction is conditioning generation on external inputs.
MusicLM supports text and melody guidance~\cite{AgostinelliEtAl2023_musiclm}, MusicGen supports text and melody conditioning in a single-stage transformer~\cite{CopetEtAl2023_musicgen}, Stable Audio Open provides latent diffusion-based text-to-audio generation~\cite{EvansEtAl2025_stable_audio_open}, and InspireMusic supports long-form generation from both text and audio prompts~\cite{ZhangEtAl2025_inspiremusic}.
Multimodal conditioning has also been explored, including image-to-music generation using valence--arousal as an intermediate representation~\cite{WangChenLi2024_image_to_music}.

\subsection{Emotion in Music Generation and Recognition}

Emotion-conditioned generation forms a more specific line within controllable music generation.
A recent survey by Dash and Agres~\cite{DashAgres2024_affective_music_generation_review} shows that affective music generation spans rule-based through deep-learning systems, while emphasizing limited labeled data and inconsistent evaluation.
In symbolic generation, ELMG conditions lyric--melody generation on positive or negative emotion~\cite{BaoSun2023_elmg}, and EmoMusicTV extends this with hierarchical Transformer-VAE modeling and emotional guidance at multiple time scales~\cite{JiYang2024_emomusictv}.
However, these studies focus on symbolic generation rather than evaluating whether modern audio-domain systems preserve target emotion.

Music emotion recognition (MER) provides the computational basis for measuring affect.
Recent work has shifted from handcrafted features toward deep and multimodal models, while continuing to face challenges in annotation and emotion representation~\cite{LiyanarachchiJoshiMeijering2025_mmer_survey}.
Emotion is commonly represented categorically~\cite{Ekman1992_basic_emotions} or dimensionally through valence and arousal~\cite{Russell1980_circumplex_affect}, with richer formulations such as the Geneva Emotional Music Scale also proposed~\cite{ZentnerGrandjeanScherer2008_gems}.
We adopt the dimensional valence--arousal framework as operationalized by Music2Emotion~\cite{KangHerremans2025_music2emotion}.

\subsection{Evaluation of Generative Music}

Current evaluation of music generation typically relies on FAD~\cite{KilgourEtAl2019_fad}, KL-based distributional measures~\cite{KullbackLeibler1951_information_sufficiency}, CLAP-based text--audio relevance~\cite{ElizaldeEtAl2023_clap}, and human quality judgments, as reported in MusicGen~\cite{CopetEtAl2023_musicgen}, Stable Audio Open~\cite{EvansEtAl2025_stable_audio_open}, and InspireMusic~\cite{ZhangEtAl2025_inspiremusic}.
These metrics assess realism, distributional similarity, and prompt adherence, but do not directly quantify whether generated music matches an intended emotional condition.
This gap motivates our evaluation: we use MER-based post-hoc analysis to examine how closely generated music follows the valence--arousal profile of the source track.

%% ================================================================
\section{Method}\label{sec:method}

\begin{figure*}[t]
\centering
\resizebox{\textwidth}{!}{%
\begin{tikzpicture}[
  >=Latex,
  font=\small,
  block/.style={rectangle, draw, rounded corners=2pt,
    minimum height=7mm, align=center, font=\small},
  data/.style={block, fill=blue!10, font=\small\bfseries},
  tool/.style={block, fill=green!12},
  build/.style={block, fill=yellow!15},
  tmod/.style={block, fill=teal!8, minimum width=22mm},
  amod/.style={block, fill=orange!12, minimum width=22mm},
  evl/.style={block, fill=red!8},
  grp/.style={draw, dashed, rounded corners=4pt, inner sep=2.5mm},
  arr/.style={->, thick},
  darr/.style={->, thick, densely dashed},
]

%% === Col 0: GTZAN ===
\node[data, text width=18mm] (gtzan) at (0,0) {%
  GTZAN\\[0.3mm]
  \normalfont\scriptsize 1{,}000 tracks\\[-0.3mm]
  \normalfont\scriptsize 10 genres%
};

%% === Col 1: Analysis (stacked) ===
\node[tool, text width=24mm] (dash) at (3.2, 1.0) {%
  \textbf{DashengLM}\\[0.2mm]
  \scriptsize Description%
};
\node[tool, text width=24mm] (m2e) at (3.2,-1.0) {%
  \textbf{Music2Emotion}\\[0.2mm]
  \scriptsize Tags + V--A%
};

%% === Col 2: Prompt Construction ===
\node[build, text width=22mm] (prompt) at (6.2, 0.2) {%
  \textbf{Prompt}\\[0.2mm]
  \scriptsize GPT-4 Rewrite%
};

%% === Col 3: Text-conditioned models (top group) ===
\node[tmod] (sa)  at (9.8, 2.0) {\scriptsize Stable Audio Open};
\node[tmod] (mgm) at (9.8, 1.1) {\scriptsize MusicGen {\tiny(medium)}};
\node[tmod] (imt) at (9.8, 0.2) {\scriptsize InspireMusic};
\node[grp, fit=(sa)(mgm)(imt),
  label={[font=\scriptsize\bfseries, anchor=south]above:Text Cond.}] (tgrp) {};

%% === Col 3: Audio-conditioned models (bottom group, extra gap for label) ===
\node[amod] (mgmel) at (9.8,-2.0) {\scriptsize MusicGen {\tiny(melody)}};
\node[amod] (ima)   at (9.8,-2.9) {\scriptsize InspireMusic};
\node[grp, fit=(mgmel)(ima),
  label={[font=\scriptsize\bfseries, anchor=south]above:Audio Cond.}] (agrp) {};

%% === Col 4: M2E on generated outputs ===
\node[tool, text width=20mm] (m2e_out) at (13.2, 0) {%
  \textbf{Music2Emotion}\\[0.2mm]
  \scriptsize Per Output%
};

%% === Col 5: Generated V-A (top) and GT V-A (bottom) ===
\node[evl, text width=20mm] (gen_va) at (16, 0.8) {%
  \textbf{Generated V--A}\\[0.2mm]
  \scriptsize Per Output%
};
\node[evl, text width=20mm] (gt) at (16, -1.2) {%
  \textbf{GT V--A}\\[0.2mm]
  \scriptsize Source Tracks%
};

%% === Col 6: Evaluation ===
\node[evl, text width=22mm, font=\small\bfseries] (eval) at (19, -0.2) {%
  Evaluation\\[0.2mm]
  \normalfont\scriptsize MAE-V, MAE-A\\[-0.3mm]
  \normalfont\scriptsize VA Dist, Corr.%
};

%% === Arrows (all use .east -> .west for horizontal midpoint alignment) ===
% GTZAN to analysis tools
\draw[arr] (gtzan.east) -- (dash.west);
\draw[arr] (gtzan.east) -- (m2e.west);

% Analysis to prompt
\draw[arr] (dash.east) -- (prompt.west);
\draw[arr] (m2e.east) -- (prompt.west);

% Prompt to text-conditioned group
\draw[arr] (prompt.east) -- (tgrp.west);

% GTZAN to audio-conditioned group (routed below)
\draw[arr, rounded corners=3pt] (gtzan.south) |- (agrp.west);

% Model groups to M2E evaluation
\draw[arr] (tgrp.east) -- (m2e_out.west);
\draw[arr] (agrp.east) -- (m2e_out.west);

% M2E evaluation output to Generated V-A
\draw[arr] (m2e_out.east) -- (gen_va.west);

% Source Music2Emotion to GT V-A (routed below audio group)
\draw[arr, rounded corners=3pt] (m2e.south) -- ++(0,-2.5) -| (gt.south);

% Both V-A blocks to Evaluation
\draw[arr] (gen_va.east) -- (eval.west);
\draw[arr] (gt.east) -- (eval.west);

\end{tikzpicture}%
}%
\caption{Overview of the evaluation pipeline. GTZAN tracks are analyzed by DashengLM and Music2Emotion to construct affect-aware text prompts. Three models generate music under text conditioning (top); two models receive the source audio directly for audio conditioning (bottom). Note that MusicGen uses different checkpoints for each mode (\texttt{medium} vs.\ \texttt{melody}). Music2Emotion is applied to all outputs to obtain generated valence--arousal estimates, which are compared against ground-truth estimates from the source tracks.}
\label{fig:pipeline}
\end{figure*}
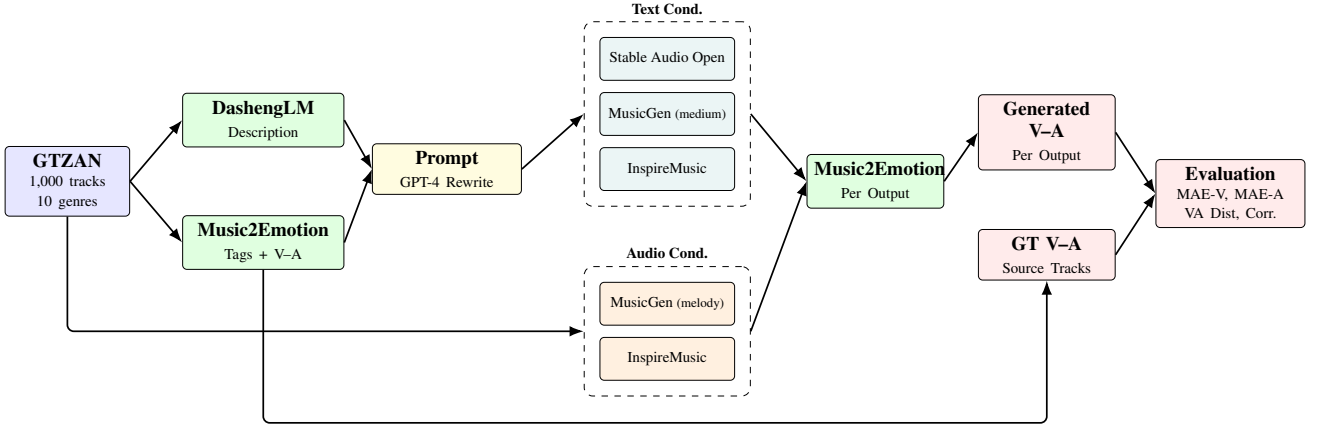

We evaluate emotion-following in generative music systems using a pipeline that extracts conditioning information from real music, generates new audio with multiple models, and compares the emotional profile of generated audio to that of the source.
Figure~\ref{fig:pipeline} provides an overview.

\subsection{Dataset}

We use GTZAN~\cite{tzanetakis2002gtzan}, which contains 1{,}000 audio tracks of 30 seconds each, spanning 10 genres (blues, classical, country, disco, hip-hop, jazz, metal, pop, reggae, rock).
All tracks are used, and genre labels are retained for genre-wise analysis.

\subsection{Audio Description and Emotion Extraction}

For each track, we obtain a semantic description using DashengLM (\texttt{mispeech/midashenglm-7b}), an audio-language model for general audio captioning~\cite{MiLMPlus2025_midashenglm}.
In parallel, we analyze the same track with Music2Emotion~\cite{KangHerremans2025_music2emotion}, which provides continuous valence--arousal predictions alongside categorical emotion tags with associated confidence scores.
Although DashengLM can produce emotion-related descriptions, we use Music2Emotion exclusively as the emotion source, ensuring a consistent valence--arousal extraction procedure across the pipeline.

\subsection{Prompt Construction}

The text prompt for each track combines the DashengLM semantic description with the top three Music2Emotion emotion tags ranked by confidence score.
These components are integrated via a rewriting step using the GPT-4 API~\cite{OpenAI2024_gpt4} to produce fluent prompts that preserve the musical description while explicitly emphasizing target affect.
For example, a source track described as a gritty electric blues piece with distorted guitar, assigned emotion tags \emph{upbeat}, \emph{funny}, and \emph{love}, yields a prompt combining these elements into a single generation instruction.
All prompts are constrained to a limited number of words to satisfy the most restrictive input constraints among the evaluated models.
All text-conditioned systems receive the same prompt for a given source track.

\subsection{Generation Models and Settings}

We evaluate three systems under five configurations:
  \begin{itemize}                                                          
      \item \textbf{Text conditioning:} Stable Audio Open                  
          (\texttt{stable-audio-open-1.0})~\cite{EvansEtAl2025_stable_audio_open},                    
          MusicGen (\texttt{musicgen-medium})~\cite{CopetEtAl2023_musicgen},                          
          and InspireMusic                                                            
          (\texttt{InspireMusic-1.5B-24kHz})~\cite{ZhangEtAl2025_inspiremusic}.                       
          Each model receives the constructed text prompt as its           
          sole input.                                                                       
      \item \textbf{Audio conditioning:} MusicGen                                    
          (\texttt{musicgen-melody})~\cite{CopetEtAl2023_musicgen}         
          and InspireMusic                                                 
          (\texttt{InspireMusic-1.5B-24kHz})~\cite{ZhangEtAl2025_inspiremusic}.                       
          Each model receives the original GTZAN track as its sole         
          input, without any text prompt. Note that MusicGen's             
          melody-conditioned checkpoint differs from the                   
          text-conditioned one (\texttt{musicgen-medium}), as it           
          conditions on chromagram features extracted from the             
          input audio. InspireMusic uses the same checkpoint for           
          both modes.                                                                       
  \end{itemize}       
Each model generates one 30-second output per track using default inference settings (no tuning of temperature, top-$k$, guidance scale, or diffusion steps).
Generated audio follows each model's native sample rate (32~kHz for MusicGen; 24~kHz for InspireMusic and Stable Audio Open) without resampling or normalization.

\subsection{Emotion-Following Metrics}\label{sec:metrics}

All generated samples are processed with Music2Emotion to obtain valence and arousal estimates in the same representation space used for the source tracks.
Let $(v_i, a_i)$ and $(\hat{v}_i, \hat{a}_i)$ denote the valence--arousal values of source track~$i$ and its generated counterpart, respectively.
We report the following metrics:
\begin{equation}
\mathrm{MAE}_v = \frac{1}{N}\sum_{i=1}^{N}|v_i - \hat{v}_i|, \quad
\mathrm{MAE}_a = \frac{1}{N}\sum_{i=1}^{N}|a_i - \hat{a}_i|
\label{eq:mae}
\end{equation}
\begin{equation}
\mathrm{Dist}_{VA}(i) = \sqrt{(v_i - \hat{v}_i)^2 + (a_i - \hat{a}_i)^2}
\label{eq:dist}
\end{equation}
where $\mathrm{MAE}_v$ and $\mathrm{MAE}_a$ are the mean absolute errors in valence and arousal, and $\mathrm{Dist}_{VA}$ is the per-track Euclidean distance in valence--arousal space.
We report the mean, standard deviation, and median of $\mathrm{Dist}_{VA}$ across tracks (denoted VA~Dist, VA~Std, and VA~Med.\ in Table~\ref{tab:main_results}).
We additionally compute Pearson correlation coefficients between source and generated valence (Corr-V) and arousal (Corr-A) to measure the degree of linear correspondence.
Lower error and distance values indicate better emotion-following; higher correlations indicate stronger correspondence with source emotion.

%% ================================================================
\section{Results}\label{sec:results}

\begin{table*}[t]
\centering
\begin{tabular}{llccccccc}
\toprule
\textbf{Model} & \textbf{Cond.} & \textbf{MAE-V}~$\downarrow$ & \textbf{MAE-A}~$\downarrow$ & \textbf{VA Dist}~$\downarrow$ & \textbf{VA Std}~$\downarrow$ & \textbf{VA Med.}~$\downarrow$ & \textbf{Corr-V}~$\uparrow$ & \textbf{Corr-A}~$\uparrow$ \\
\midrule
Stable Audio  & text  & 0.821 & 1.384 & 1.714 & 0.909 & 1.559 & 0.465 & 0.659 \\
InspireMusic  & text  & 0.825 & \textbf{0.953} & 1.367 & 0.881 & \textbf{1.150} & 0.441 & 0.613 \\
MusicGen (medium)     & text  & \textbf{0.744} & 0.983 & \textbf{1.362} & \textbf{0.704} & 1.276 & \textbf{0.519} & \textbf{0.666} \\
\midrule
InspireMusic  & audio & 1.223 & 1.563 & 2.101 & 1.258 & 1.895 & 0.031 & $-$0.017 \\
MusicGen (melody)     & audio & 0.966 & 1.416 & 1.804 & 1.068 & 1.618 & 0.254 & 0.311 \\
\bottomrule
\end{tabular}
\caption{Emotion-following results across models and conditioning types. MAE-V/A: mean absolute error in valence/arousal. VA Dist/Std/Med.: mean, standard deviation, and median of per-track Euclidean distance in valence--arousal space. Corr-V/A: Pearson correlation with source valence/arousal. Best overall values are \textbf{bolded}. A horizontal rule separates text- from audio-conditioned systems.}
\label{tab:main_results}
\end{table*}

\begin{figure*}[b]
    \centering
    \includegraphics[width=\textwidth]{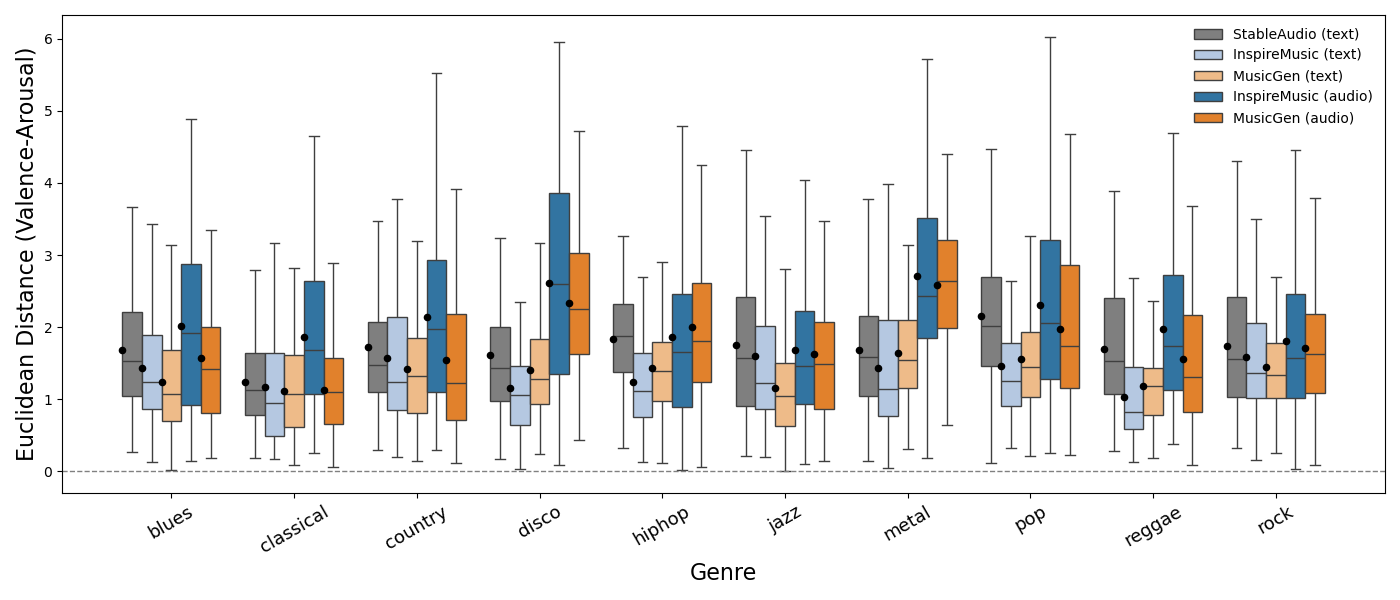}
    \caption{Distribution of valence--arousal Euclidean distance across the 10 GTZAN genres. Lower values indicate better emotion alignment. Each box shows the interquartile range with median (line) and mean (dot). Text-conditioned models (lighter shades) are shown alongside audio-conditioned models (darker shades).}
    \label{fig:boxplot}
\end{figure*}

\subsection{Overall Emotion-Following Performance}

Table~\ref{tab:main_results} summarizes the overall results.
Two main findings emerge.

\textbf{Text conditioning outperforms audio conditioning.}
The two best systems are MusicGen~(text) and InspireMusic~(text), with mean VA distances of 1.362 and 1.367, respectively.
Stable Audio Open~(text) achieves a higher distance of 1.714 but still outperforms both audio-conditioned systems.
MusicGen~(audio) reaches 1.804, while InspireMusic~(audio) performs worst at 2.101.
The lower standard deviations for text-conditioned models further indicate more consistent emotion alignment across tracks.

\textbf{Valence is preserved more reliably than arousal.}
Across all five systems, MAE-V is consistently lower than MAE-A.
MusicGen~(text) achieves the lowest valence error (0.744), while InspireMusic~(text) achieves the lowest arousal error (0.953).
The systematic gap suggests that dynamic intensity and activation level are harder for current generation systems to match than the positive--negative emotional character.

The correlation analysis confirms these patterns.
MusicGen~(text) achieves the strongest correspondence with source emotion in both dimensions (Corr-V = 0.519, Corr-A = 0.666).
Audio-conditioned InspireMusic shows near-zero correlation (Corr-V = 0.031, Corr-A = $-$0.017), indicating that its audio conditioning does not reliably preserve the source emotional profile.

\subsection{Genre-Wise Analysis}

Figure~\ref{fig:boxplot} shows the distribution of VA distances across the 10 GTZAN genres.
Several patterns emerge.

First, the advantage of text conditioning is consistent across genres: text-conditioned MusicGen and InspireMusic exhibit lower medians and tighter interquartile ranges than their audio-conditioned counterparts in nearly every genre.

Second, genre difficulty varies substantially.
Genres such as \textbf{classical} and \textbf{country} yield lower distances and tighter spreads, suggesting their emotional characteristics are more reliably reproduced.
In contrast, \textbf{disco}, \textbf{metal}, and \textbf{pop} show larger distances and broader distributions, indicating greater difficulty in emotion preservation.
This may reflect that genres with high-energy, rhythmically driven affective profiles are harder to match, particularly when arousal is the dominant emotional dimension.

Third, audio-conditioned systems are markedly less stable.
InspireMusic~(audio) in particular exhibits large spread and high upper-range values across several genres, confirming that direct audio conditioning does not automatically preserve source emotion and may introduce greater affective variance.

\subsection{Text vs.\ Audio Conditioning}

\begin{table}[t]
\centering
\small
\begin{tabular}{@{}lcrr@{}}
\toprule
\textbf{Comparison} & \textbf{Win~\%} & \textbf{$p$-value} & \textbf{$|d|$} \\
\midrule
InspireMusic: text vs.\ audio & 67.3 & $<$0.001 & 0.502 \\
MusicGen: text vs.\ audio     & 64.5 & $<$0.001 & 0.411 \\
\bottomrule
\end{tabular}
\caption{Paired Wilcoxon signed-rank tests comparing text vs.\ audio conditioning within each model family on VA distance. Win~\% indicates how often text conditioning achieves lower distance. $|d|$: absolute Cohen's~$d$ effect size.}
\label{tab:statistical_results}
\end{table}

Table~\ref{tab:statistical_results} provides within-family statistical comparisons using paired Wilcoxon signed-rank tests on the per-track VA distances.
Text conditioning significantly outperforms audio conditioning for both InspireMusic (win rate 67.3\%, $p < 0.001$, $|d| = 0.502$) and MusicGen (win rate 64.5\%, $p < 0.001$, $|d| = 0.411$), with moderate effect sizes in both cases.

This result is noteworthy because one might expect audio conditioning to better preserve the source emotion, since the model receives the actual audio.
Our results suggest the opposite: explicit text prompts combining semantic description with emotion tags are more effective for preserving target affect than source-audio conditioning alone.

Importantly, the two model families differ in how cleanly they isolate the effect of conditioning mode.
InspireMusic uses the same checkpoint for both text and audio conditioning, so its comparison (67.3\% win rate, $|d| = 0.502$) directly reflects the effect of conditioning type.
For MusicGen, however, text and audio conditioning rely on different checkpoints (\texttt{musicgen-medium} vs.\ \texttt{musicgen-melody}), which differ in both training and conditioning pathway; its comparison therefore conflates conditioning mode with model variant.
The stronger and cleaner InspireMusic result thus provides the more controlled evidence that explicit text-based emotion cues outperform audio conditioning for emotion preservation.

\subsection{Source--Generated Correlation}

\begin{figure*}[t]
\centering
\includegraphics[width=\textwidth]{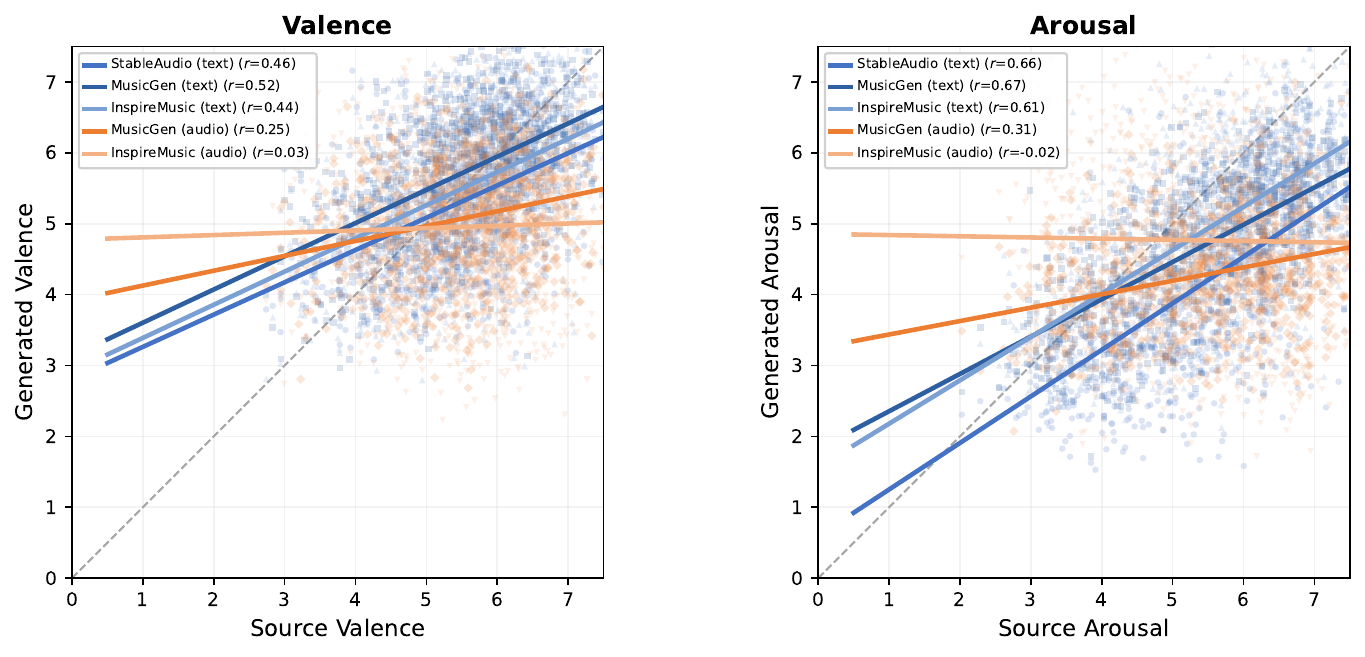}
\caption{Source vs.\ generated valence (left) and arousal (right) for all five systems. Each point represents one track; regression lines and Pearson $r$ values are shown per model. The dashed diagonal indicates perfect emotion preservation. Text-conditioned models (blue) exhibit steeper slopes closer to the diagonal than audio-conditioned models (orange), and the valence panel shows stronger correlations overall than the arousal panel.}
\label{fig:correlation_va}
\end{figure*}

Figure~\ref{fig:correlation_va} visualizes the relationship between source and generated emotion values across all five systems.
In the valence panel, text-conditioned models produce regression lines with steeper slopes and higher correlations ($r = 0.44$--$0.52$), while audio-conditioned MusicGen achieves a weaker correspondence ($r = 0.25$) and InspireMusic~(audio) shows near-zero correlation ($r = 0.03$).
The arousal panel reveals a similar pattern, with text-conditioned models reaching $r = 0.61$--$0.67$ and audio-conditioned models falling substantially below.
Notably, the regression lines for audio-conditioned models are considerably flatter, indicating that these systems tend to generate outputs with a narrower range of valence and arousal values regardless of the source emotion.
This visualization complements the aggregate metrics in Table~\ref{tab:main_results} by revealing that the emotion-following gap between text and audio conditioning is not driven by outliers but reflects a systematic difference across the full range of source emotion values.

%% ================================================================
\section{Discussion}\label{sec:discussion}

Our experiments yield three main findings that we discuss in turn.

\textbf{Text conditioning is more effective than audio conditioning for emotion-following.}
This finding is counterintuitive: one might expect that providing the source audio directly would better preserve its emotional character.
We hypothesize that text prompts provide explicit, parameterizable affective information through emotion labels, which generation models can directly attend to during synthesis.
In contrast, audio conditioning requires implicit extraction and preservation of emotional attributes from the source signal, which may compete with other perceptual features such as timbre and rhythm during the generation process.

\textbf{Valence is preserved more reliably than arousal.}
This asymmetry may reflect how musical attributes map to these affective dimensions.
Valence is strongly associated with mode and harmonic content---attributes that are relatively straightforward to convey through text descriptions.
Arousal relates more to tempo, dynamics, and energy, which may be harder to control precisely through text prompts and more sensitive to architectural differences across generation systems.

\textbf{Emotion-following varies by genre.}
Genres with distinctive high-energy affective profiles (disco, metal, pop) pose greater challenges than genres with more moderate or consistent emotional characteristics (classical, country).
This suggests that current generation models may default toward a moderate affective range, making extreme or genre-specific emotional profiles harder to reproduce.

A practical implication of these findings is that incorporating explicit emotion labels into text prompts is currently a more reliable strategy than audio-to-audio conditioning for developers building emotion-aware music generation systems.

\textbf{Limitations.}
Several caveats apply.
First, we use Music2Emotion as both the ground-truth emotion source and the evaluation model, meaning our metrics reflect consistency within a single MER system's representation space rather than perceptual ground truth.
Future work should incorporate human listener evaluations of emotional congruence.
Second, GTZAN, while widely used, has known limitations including potential artist repetition and label ambiguities~\cite{Sturm2013_gtzan_analysis}; validation on larger and more diverse corpora would strengthen generalizability.
Third, we generate a single output per source track with default inference settings; multiple generations and optimized parameters might reveal different performance characteristics.
Additionally, for MusicGen, text and audio conditioning use different checkpoints with distinct training procedures, so the within-family comparison conflates conditioning mode with model differences; InspireMusic, which uses the same checkpoint for both modes, provides the cleaner comparison.
Finally, GPT-4-based prompt rewriting introduces a dependency on language model quality, and alternative rewriting strategies could yield different results.

%% ================================================================
\section{Conclusion}\label{sec:conclusion}

We presented a systematic evaluation pipeline for measuring emotion-following in generative music, comparing three models under five conditioning configurations across 1{,}000 tracks.
Our pipeline, combining audio captioning, music emotion recognition, and generation, provides a reproducible framework for assessing affective controllability.
The finding that text conditioning with explicit emotion cues significantly outperforms audio conditioning ($p < 0.001$) highlights the importance of explicit affective specification in prompts.
The consistent gap between valence and arousal preservation, and the substantial genre-dependent variation, point to specific directions for improving emotion-aware generation.
Future work should extend this evaluation to additional models, larger datasets, human perceptual validation, and fine-grained analysis of how musical attributes mediate emotion preservation.

%% ================================================================
% References
\bibliography{ISMIRtemplate}

@article{WangEtAl2024_review_music_generation,
  author  = {Wang, Lei and Zhao, Ziyi and Liu, Hanwei and Pang, Junwei and Qin, Yi and Wu, Qidi},
  title   = {A review of intelligent music generation systems},
  journal = {Neural Computing and Applications},
  year    = {2024},
  volume  = {36},
  pages   = {6381--6401},
  doi     = {10.1007/s00521-024-09418-2}
}

@article{AgostinelliEtAl2023_musiclm,
  author        = {Agostinelli, Andrea and Denk, Timo I. and Borsos, Zal{\'a}n and Engel, Jesse and Verzetti, Mauro and Caillon, Antoine and Huang, Qingqing and Jansen, Aren and Roberts, Adam and Tagliasacchi, Marco and Sharifi, Matt and Zeghidour, Neil and Frank, Christian},
  title         = {{MusicLM}: Generating Music From Text},
  journal       = {arXiv preprint arXiv:2301.11325},
  year          = {2023},
  eprint        = {2301.11325},
  archivePrefix = {arXiv},
  primaryClass  = {cs.SD}
}

@inproceedings{CopetEtAl2023_musicgen,
  author    = {Copet, Jade and Kreuk, Felix and Gat, Itai and Remez, Tal and Kant, David and Synnaeve, Gabriel and Adi, Yossi and D{\'e}fossez, Alexandre},
  title     = {Simple and Controllable Music Generation},
  booktitle = {Advances in Neural Information Processing Systems},
  volume    = {36},
  pages     = {47704--47720},
  year      = {2023}
}

@inproceedings{EvansEtAl2025_stable_audio_open,
  author    = {Evans, Zach and Parker, Julian D. and Carr, CJ and Zukowski, Zack and Taylor, Josiah and Pons, Jordi},
  title     = {Stable Audio Open},
  booktitle = {ICASSP 2025 -- 2025 IEEE International Conference on Acoustics, Speech and Signal Processing (ICASSP)},
  year      = {2025},
  pages     = {1--5},
  doi       = {10.1109/ICASSP49660.2025.10888461}
}

@article{ZhangEtAl2025_inspiremusic,
  author        = {Zhang, Chong and Ma, Yukun and Chen, Qian and Wang, Wen and Zhao, Shengkui and Pan, Zexu and Wang, Hao and Ni, Chongjia and Nguyen, Trung Hieu and Zhou, Kun and Jiang, Yidi and Tan, Chaohong and Gao, Zhifu and Du, Zhihao and Ma, Bin},
  title         = {{InspireMusic}: Integrating Super Resolution and Large Language Model for High-Fidelity Long-Form Music Generation},
  journal       = {arXiv preprint arXiv:2503.00084},
  year          = {2025},
  eprint        = {2503.00084},
  archivePrefix = {arXiv},
  primaryClass  = {cs.SD}
}

@article{DhariwalEtAl2020_jukebox,
  author  = {Dhariwal, Prafulla and Jun, Heewoo and Payne, Christine and Kim, Jong Wook and Radford, Alec and Sutskever, Ilya},
  title   = {Jukebox: A Generative Model for Music},
  journal = {arXiv preprint arXiv:2005.00341},
  year    = {2020}
}

@inproceedings{HuangEtAl2019_music_transformer,
  author    = {Huang, Cheng-Zhi Anna and Vaswani, Ashish and Uszkoreit, Jakob and Simon, Ian and Hawthorne, Curtis and Shazeer, Noam and Dai, Andrew M. and Hoffman, Matthew D. and Dinculescu, Monica and Eck, Douglas},
  title     = {Music Transformer: Generating Music with Long-Term Structure},
  booktitle = {Proceedings of the 7th International Conference on Learning Representations (ICLR)},
  year      = {2019}
}

@inproceedings{Conklin2003_statistical,
  author    = {Conklin, Darrell},
  title     = {Music Generation from Statistical Models},
  booktitle = {Proceedings of the AISB 2003 Symposium on Artificial Intelligence and Creativity in the Arts and Sciences},
  address   = {Aberystwyth, Wales},
  year      = {2003},
  pages     = {30--35}
}

@inproceedings{YangEtAl2017_midinet,
  author    = {Yang, Li-Chia and Chou, Szu-Yu and Yang, Yi-Hsuan},
  title     = {{MidiNet}: A Convolutional Generative Adversarial Network for Symbolic-Domain Music Generation},
  booktitle = {Proceedings of the 18th International Society for Music Information Retrieval Conference (ISMIR)},
  year      = {2017},
  pages     = {324--331}
}

@inproceedings{MittalEtAl2021_symbolic_diffusion,
  author    = {Mittal, Gautam and Engel, Jesse and Hawthorne, Curtis and Simon, Ian},
  title     = {Symbolic Music Generation with Diffusion Models},
  booktitle = {Proceedings of the 22nd International Society for Music Information Retrieval Conference (ISMIR)},
  year      = {2021}
}

@article{DashAgres2024_affective_music_generation_review,
  author  = {Dash, Adyasha and Agres, Kathleen},
  title   = {{AI}-Based Affective Music Generation Systems: A Review of Methods and Challenges},
  journal = {ACM Computing Surveys},
  year    = {2024},
  volume  = {56},
  number  = {11},
  doi     = {10.1145/3672554}
}

@article{LiyanarachchiJoshiMeijering2025_mmer_survey,
  author  = {Liyanarachchi, Rashini and Joshi, Aditya and Meijering, Erik},
  title   = {A Survey on Multimodal Music Emotion Recognition},
  journal = {arXiv preprint arXiv:2504.18799},
  year    = {2025}
}

@article{KangHerremans2025_music2emotion,
  author  = {Kang, Jaeyong and Herremans, Dorien},
  title   = {Towards Unified Music Emotion Recognition across Dimensional and Categorical Models},
  journal = {arXiv preprint arXiv:2502.03979},
  year    = {2025}
}

@article{BaoSun2023_elmg,
  author  = {Bao, Chunhui and Sun, Qianru},
  title   = {Generating Music With Emotions},
  journal = {IEEE Transactions on Multimedia},
  year    = {2023},
  volume  = {25},
  pages   = {3602--3614},
  doi     = {10.1109/TMM.2022.3163543}
}

@article{JiYang2024_emomusictv,
  author  = {Ji, Shulei and Yang, Xinyu},
  title   = {{EmoMusicTV}: Emotion-Conditioned Symbolic Music Generation With Hierarchical Transformer {VAE}},
  journal = {IEEE Transactions on Multimedia},
  year    = {2024},
  volume  = {26},
  pages   = {1076--1088},
  doi     = {10.1109/TMM.2023.3276177}
}

@article{WangChenLi2024_image_to_music,
  author  = {Wang, Yajie and Chen, Mulin and Li, Xuelong},
  title   = {Continuous Emotion-Based Image-to-Music Generation},
  journal = {IEEE Transactions on Multimedia},
  year    = {2024},
  volume  = {26},
  pages   = {5670--5679},
  doi     = {10.1109/TMM.2023.3338089}
}

@article{Russell1980_circumplex_affect,
  author  = {Russell, James A.},
  title   = {A Circumplex Model of Affect},
  journal = {Journal of Personality and Social Psychology},
  year    = {1980},
  volume  = {39},
  number  = {6},
  pages   = {1161--1178},
  doi     = {10.1037/h0077714}
}

@article{Ekman1992_basic_emotions,
  author  = {Ekman, Paul},
  title   = {An Argument for Basic Emotions},
  journal = {Cognition and Emotion},
  year    = {1992},
  volume  = {6},
  number  = {3--4},
  pages   = {169--200},
  doi     = {10.1080/02699939208411068}
}

@article{ZentnerGrandjeanScherer2008_gems,
  author  = {Zentner, Marcel and Grandjean, Didier and Scherer, Klaus R.},
  title   = {Emotions Evoked by the Sound of Music: Characterization, Classification, and Measurement},
  journal = {Emotion},
  year    = {2008},
  volume  = {8},
  number  = {4},
  pages   = {494--521},
  doi     = {10.1037/1528-3542.8.4.494}
}

@article{KilgourEtAl2019_fad,
  author  = {Kilgour, Kevin and Zuluaga, Mauricio and Roblek, Dominik and Sharifi, Matthew},
  title   = {{Fr{\'e}chet Audio Distance}: A Metric for Evaluating Music Enhancement Algorithms},
  journal = {arXiv preprint arXiv:1812.08466},
  year    = {2019}
}

@article{KullbackLeibler1951_information_sufficiency,
  author  = {Kullback, S. and Leibler, R. A.},
  title   = {On Information and Sufficiency},
  journal = {The Annals of Mathematical Statistics},
  year    = {1951},
  volume  = {22},
  number  = {1},
  pages   = {79--86}
}

@inproceedings{ElizaldeEtAl2023_clap,
  author    = {Elizalde, Benjamin and Deshmukh, Soham and Al Ismail, Mahmoud and Wang, Huaming},
  title     = {{CLAP}: Learning Audio Concepts From Natural Language Supervision},
  booktitle = {ICASSP 2023 -- 2023 IEEE International Conference on Acoustics, Speech and Signal Processing (ICASSP)},
  year      = {2023},
  pages     = {1--5},
  doi       = {10.1109/ICASSP49357.2023.10095889}
}

@article{MiLMPlus2025_midashenglm,
  author  = {{MiLM Plus}},
  title   = {{MiDashengLM}: Efficient Audio Understanding with General Audio Captions},
  journal = {arXiv preprint arXiv:2508.03983},
  year    = {2025}
}

@article{OpenAI2024_gpt4,
  author  = {{OpenAI}},
  title   = {{GPT}-4 Technical Report},
  journal = {arXiv preprint arXiv:2303.08774},
  year    = {2024}
}

@article{tzanetakis2002gtzan,
  author  = {Tzanetakis, George and Cook, Perry},
  title   = {Musical Genre Classification of Audio Signals},
  journal = {IEEE Transactions on Speech and Audio Processing},
  volume  = {10},
  number  = {5},
  pages   = {293--302},
  year    = {2002},
  doi     = {10.1109/TSA.2002.800560}
}

@article{Sturm2013_gtzan_analysis,
  author  = {Sturm, Bob L.},
  title   = {The {GTZAN} Dataset: Its Contents, Its Faults, Their Effects on Evaluation, and Its Future Use},
  journal = {arXiv preprint arXiv:1306.1461},
  year    = {2013}
}

\end{document}